\documentclass[11pt,a4paper]{article}
\usepackage[utf8]{inputenc}
\usepackage{amsmath}
\usepackage{amsfonts}
\usepackage{amssymb}
\usepackage{tikz} 
\usepackage{float} 
\newcommand{\dt}[1][t]{\,\mathrm{d}{#1}}	
\newcommand{\bracket}[1]{\left( {#1} \right)}	
\newcommand{\der}[2][]{\frac{\mathrm{d} {#1} }{\mathrm{d}{#2}}}
\newcommand{\sder}[2][]{\frac{\mathrm{d}^2 {#1} }{\mathrm{d}{#2}^2}}
\newcommand{\partialder}[2][]{\frac{\partial {#1} }{\partial{#2}}}

\newcommand{\abs}[1]{\left| {#1} \right|} 		
\newcommand{\lagr}{\mathcal{L}}
\newcommand{\ham}{\mathcal{H}}

\let\oldref\ref
\renewcommand{\ref}[1]{(\oldref{#1})}
\renewcommand{\epsilon}{\varepsilon}
\let\oldint\int
\renewcommand{\int}{\oldint\limits}

\begin{document}
\title{Coupling constant metamorphosis, Fermat principle and light propagation in Kerr metric}
\author{Joanna Piwnik\thanks{joanna.piwnik@edu.uni.lodz.pl}, Cezary Gonera\thanks{cezary.gonera@uni.lodz.pl}, Joanna Gonera\thanks{joanna.gonera@uni.lodz.pl}, Piotr Kosiński\thanks{piotr.kosinski@uni.lodz.pl}}
\date{}
\maketitle
\begin{center}
Faculty of Physics and Applied Informatics\\
University of Lodz, Lodz, Poland
\end{center} 

\begin{abstract}
The geodesics of Kerr's metric are described by the four-dimensional Hamiltonian dynamics integrable in the Arnold-Liouville sense. It can be reduced to two-dimensional one by the use of Fermat's principle. The resulting Hamiltonian is, however, rather complicated. We show how one can apply the coupling constant metamorphosis to simplify the Hamiltonian to the one quadratic in momenta and depending on the initial "energy" as parameter. It describes a simple dynamics of two non-linear oscillators and can be integrated directly or evaluated in the framework of perturbation theory by adopting the elegant Lindstedt--Poincar\'e algorithm. The idea of coupling constant metamorphosis is also applied to the Myers--Perry metric --- a five dimensional generalization of Kerr's metric. The case of single rotation parameter is considered in some detail.
\end{abstract}

\section{Introduction}
An interesting method for studying dynamical systems has been proposed by Hietarinta et al. \cite{1} under the name of coupling-constant metamorphosis. It allows us to relate various Hamiltonians sharing the same (unparametrized) trajectories in phase space. This method has been considerably generalized in the recent paper \cite{2}. We apply it here to the study of light rays propagation in Kerr's metric. We show how its use leads to the considerable simplification providing a nice picture of two-dimensional integrable dynamics.  We also outline the application to the study of light propagation in 5d Myers--Perry space-time which provides the generalization of Kerr's metric to higher dimensions.
 
In general relativity the light propagation is described by the solutions to Maxwell's equations in curved space-time. In the short wavelength limit they lead to the geometric optic picture with its notion of light rays. The ray trajectories are the null geodesics which, in turn, are described by Euler--Lagrange equations resulting from the quadratic (in generalized velocities) Lagrangian, with the affine parameter as the evolution one. Consequently, one can apply the whole machinery of analytical mechanics to describe the light rays trajectories. In particular, one can ask if the relevant dynamics is integrable in the Arnold--Liouville sense. Since we are dealing with the dynamical system with four degrees of freedom, four Poisson-commuting integrals of motion are needed. One can expect that the more symmetric the background metric is, the more integrals of motion are expected to exist. The Kerr metric is time-independent and axially symmetric; consequently the generalized momenta $\pi_t$ and $\pi_\phi$ are integrals of motion. Moreover, due to the fact that the metric does not depend on the affine parameter, the Lagrangian (numerically equal to the Hamiltonian) is an integral of motion. So we have three mutually commuting integrals of motion and only one is lacking. The remaining integral has been found by Carter \cite{3}. It is not related to any Killing vector but rather ascribed to the existence of Killing tensor \cite{5}. From the point of view of analytical mechanics this implies that the Carter integral of motion results from the symmetry transformations which are canonical transformations not reducing to the point ones.
 
Once the Arnold--Liouville integrability is established, the relevant geodesic equations can be solved in quadratures which results in description in terms of elliptic functions \cite{6,7,8}.
 
It should be noted, however, that the integrals of motion discussed above do not have direct physical meaning. In fact, they depend on the actual choice of affine parameter, which is defined up to an affine transformation.
 
Alternative description of light rays propagation is provided by Fermat's principle \cite{9,10,11,12,13,14,15,16,17,18,19}. It takes a particularly simple form in the case of a constant gravitational field, both static and stationary \cite{14}. One solves the equation
\begin{equation} 
\label{-1}
\bracket{\der[s]{\sigma}}^2 = 0
\end{equation}
with respect to $\der[x^0]{\sigma}$ (both solutions can be used):
\begin{equation}
\label{0}
\der[x^0]{\sigma} = \lagr \bracket{\vec{x}, \der[\vec{x}]{\sigma}}
\end{equation}
and considers the function $\lagr$ as the Lagrangian generating the Euler--Lagrange equations describing the light rays trajectories in three-dimensional submanifolds of constant time $x^0$. An important property of dynamics described by $\lagr$ is its reparametrization invariance. We don't have to use $\sigma$ as an evolution parameter any longer, while $\lagr$ is degenerate. In order to get a regular dynamics one should choose a gauge. If some of the variables $x^i$, $i=1,2,3$, say $x^3$, is cyclic, the convenient choice is $\sigma=x^3$. Then we are left with two degrees of freedom, $x^1$ and $x^2$. It is then straightforward to compute the Hamiltonian and write out the canonical equations of motion. The simplification consists in reduction of the number of degrees of freedom from four to two. There is, however, some price to be paid for this simplification. The resulting Hamiltonian is quite complicated being the nonpolynomial function of momenta. At this point the coupling constant metamorphosis comes into play. It follows from eq. \ref{33} below that the Hamiltonian obeys a quadratic equation with coefficients depending on phase space variables. The method sketched in Sec. II allows to replace the initial Hamiltonian by the simpler one, quadratic in momenta, which depends on initial energy as a parameter. The family of trajectories with varying energies is then described by the family of new simple Hamiltonians depending on one parameter; all trajectories correspond to vanishing new energy. The new form of dynamics has two advantages: it is explicitly Arnold--Liouville integrable with the additional integral of motion (a counterpart of Carter's constant) being independent of the choice of affine parameter; after a simple redefinition of evolution parameter (a counterpart of Mino time \cite{20}) the equations of motion (actually, already once integrated) completely decouple. As a result one obtains two decoupled nonlinear oscillators. They can be immediately integrated, by separation of variables, in terms of elliptic functions. Moreover, in the weak deflection limit one deals with small nonlinear oscillations which can be systematically described in terms of Lindstedt--Poincar\'e expansion \cite{21,22,23}.
 
The paper is organized as follows. In Sec. 2 we remind the particular case of the method developed in \cite{2}. It is then applied in Sec.3 to the Fermat principle for time independent gravitational field possessing an additional cyclic coordinate. The integrability of null geodesics in Kerr metric is described in Sec. 4. Sec. 5 is devoted to the analysis of weak deflection limit. In Sec. 6 we consider the Myers--Perry metric in five dimensions. We show that the method outlined in previous sections works also in the case with no essential modifications. The case of single rotation parameter is considered in some detail. Finally, Sec. 7 contains some final remarks.

\section{Coupling constant metamorphosis}

The method called coupling constant metamorphosis has been proposed in \cite{1} and considerably extended in \cite{2}. Here we describe briefly its particular form we shall use in the following sections.
 
Assume we have a Hamiltonian system with $f$ degrees of freedom described by the Hamiltonian
\begin{equation}
\label{1}
\tilde{H} = \tilde{H} \bracket{\underline{q}, \underline{p}},
\end{equation}
where $\underline{q} \equiv \bracket{q_1, ..., q_f}$, $ \underline{p} = \bracket{p_1, ..., p_f}$. Assume further that $\tilde{H}$ obeys a functional equation of the form:
\begin{equation}
\label{2}
F \bracket{\underline{q}, \underline{p}; \tilde{H}\bracket{\underline{q}, \underline{p}}} \equiv 0,
\end{equation} 
where
\begin{equation}
\label{3}
\partialder[F]{\tilde{H}} \not\equiv 0.
\end{equation}
Define the family of Hamiltonians depending on additional parameter $\lambda$:
\begin{equation}
\label{4}
H = F \bracket{\underline{q}, \underline{p}; \lambda}.
\end{equation}
It is then straightforward to derive the following result \cite{2}. Let $\underline{q} = \underline{q}\bracket{\tilde{t}}$, $\underline{p} = \underline{p}\bracket{\tilde{t}}$ be a solution to the canonical equations:
\begin{align}
\label{5}
\der[q_i]{\tilde{t}} &= \partialder[\tilde{H}]{p_i}
\\
\label{6}
\der[p_i]{\tilde{t}} &= - \partialder[\tilde{H}]{q_i},
\end{align}
carrying the energy $\tilde{H}=\tilde{E}$. Define new evolution parameter $t$ by
\begin{equation}
\label{7}
\der[\tilde{t}]{t} = - \partialder[F \bracket{\underline{q}\bracket{\tilde{t}}, \underline{p}\bracket{\tilde{t}}; \tilde{E}}]{\tilde{E}}.
\end{equation}
Then $\underline{q} = \underline{q}\bracket{\tilde{t}\bracket{t}}$, $\underline{p} = \underline{p}\bracket{\tilde{t}\bracket{t}}$ are the solutions to canonical equations
\begin{align}
\label{8}
\der[q_i]{t} &= \partialder[H \bracket{\underline{q}, \underline{p}; \tilde{E}}]{p_i}
\\
\label{9}
\der[p_i]{t} &= - \partialder[H \bracket{\underline{q}, \underline{p}; \tilde{E}}]{q_i}
\end{align}
carrying vanishing energy $H=E=0$. In other words, the family of unparameterized trajectories of $\tilde{H}$, corresponding to different energies $\tilde{E}$, coincides with the family of those described by $\left. H\bracket{\lambda}\right|_{\lambda = \tilde{E}}$ and carrying the vanishing energy.
\\
Let us conclude with an important remark. Given a function
\begin{equation}
\label{10}
G = G \bracket{\underline{q}, \underline{p}} \not\equiv 0
\end{equation}
one can replace eq. \ref{2} with
\begin{equation}
\bracket{GF}\bracket{\underline{q}, \underline{p}; \tilde{H}} \equiv 0,
\end{equation}
defining the same Hamiltonian $\tilde{H} = \tilde{H} \bracket{\underline{q}, \underline{p}}$. Obviously, some care must be exercised on submanifolds $G\bracket{\underline{q}, \underline{p}} = 0$.
 
\section{Fermat principle and coupling constant metamorphosis}

We shall apply the formalism described above to the problem of light propagation in the gravitational field. In the short wavelength limit one can refer to the geometric optics and Fermat's principle. The latter takes a particularly simple form in the case of constant gravitational field. Its very nice derivation, based directly on equivalence principle, has been given in \cite{14}. However, we prefer to follow somewhat more formal path \cite{19}. The geodesic equation can be derived from the Lagrangian
\begin{equation}
\label{12}
L = \frac{1}{2} g_{\mu \nu} \bracket{x} \der[x^\mu]{\sigma} \der[x^\nu]{\sigma}
\end{equation}
where $\sigma$ is an affine parameter. The light trajectories are singled out by the null geodesic condition
\begin{equation}
\label{13}
g_{\mu\nu}\bracket{x} \der[x^\mu]{\sigma} \der[x^\nu]{\sigma} = 0
\end{equation}
Assume now that we are dealing with constant gravitational field in appropriate coordinates, i.e. $g_{\mu\nu}\bracket{x}$ does not depend on $x^0$. Then, solving \ref{13} with respect to $\der[x^0]{\sigma}$ one obtains
\begin{equation}
\label{14}
\der[x^0]{\sigma} \equiv \der[x^0]{\sigma}\bracket{\vec{x}, \der[\vec{x}]{\sigma}} \equiv \lagr\bracket{\vec{x}, \der[\vec{x}]{\sigma}},\hspace{3mm} \vec{x} \equiv \bracket{x^1, x^2, x^3}.
\end{equation}  
It can be shown \cite{19} that the trajectories resulting from the Lagrange equations for $\lagr$ are the 3-space projections of null geodesics described by $L$. Moreover, due to eq. \ref{14}, $x^0$ plays here the role of action. 
\\
Explicitly, the solution to eq. \ref{13} reads (cf. also \cite{14}): 
\begin{equation}
\label{15}
\lagr \equiv \der[x^0]{\sigma} = - \frac{g_{0i}}{g_{00}} \der[x^i]{\sigma} + \frac{1}{\sqrt{g_{00}}} \der[l]{\sigma},
\end{equation}
where
\begin{equation}
\label{16}
\dt[l]^2 \equiv \bracket{-g_{ij} + \frac{g_{0i}g_{0j}}{g_{00}}} \dt[x]^i \dt[x]^j \equiv \gamma_{ij} \dt[x]^i \dt[x]^j  
\end{equation} 
denotes the spatial metric \cite{14}. Actually, there are two solutions to eq. \ref{13} and, in the defined sense, both can be used. However, this is not important for our purposes, as we shall see later. 

Concluding, the Fermat principle for constant gravitational field can be put in the Lagrangian form with $\lagr$, eq. \ref{14}, playing the role of the relevant Lagrangian and $x^0$ being the action variable.

However, what we really need in order to apply the algorithm of Sec. II, is the Hamiltonian formalism. The trouble is that the Lagrangian dynamics, defined by the Lagrangian \ref{15}, is reparametrization invariant and, therefore, degenerate. In order to construct the Hamiltonian formalism one has either to use the Dirac approach to constrained systems or impose the gauge condition. In view of application to the Kerr metric we will be interested in the case when, apart from $x^0$, some other coordinate, say $x^3$ (actually, the azimuthal angle), is cyclic. A convenient gauge choice is then
\begin{equation}
\label{17}
x^3 - \sigma = 0
\end{equation} 
Replacing $\sigma$ by $x^3$ in eq. \ref{15} one obtains nondegenerate Lagrangian which allows passing to the Hamiltonian formalism by direct application of Legendre transformation. This is straightforward but slightly tedious. Moreover, an alternative procedure is more elegant and suitable for applying the algorithm described in the previous section. To begin with, note that the Lagrangian \ref{12} is nondegenerate. The relevant momenta read 
\begin{equation}
\label{18}
\pi_\mu \equiv \partialder[L]{\bracket{\der[x^\mu]{\sigma}}} = g_{\mu\nu}\bracket{x} \der[x^\nu]{\sigma},
\end{equation}
 while the Hamiltonian, numerically equal to the Lagrangian $L$ takes the form 
\begin{equation}
\label{19}
H = \frac{1}{2} g^{\mu\nu}\bracket{x} \pi_\mu \pi_\nu.
\end{equation} 
The constraint \ref{13} implies 
\begin{equation}
\label{20}
g^{\mu\nu}\bracket{x} \pi_\mu \pi_\nu = 0.
\end{equation}
It is easy to relate the momenta $\pi_\mu$ to those resulting from $\lagr$, 
\begin{equation}
\label{21}
p_i = \partialder[\lagr]{\bracket{\der[x^i]{\sigma}}}
\end{equation}  
In fact, eqs. \ref{13}, \ref{14}, yield 
\begin{equation}
L\bracket{\lagr\bracket{\vec{x}, \der[\vec{x}]{\sigma}}, \vec{x}, \der[\vec{x}]{\sigma}} \equiv 0.
\end{equation}  
Taking derivative with respect to $\der[\vec{x}]{\sigma}$ one finds
\begin{equation}
\label{23}
\partialder[L]{\bracket{\der[\vec{x}]{\sigma}}} +\partialder[L]{\bracket{\der[x^0]{\sigma}}} \partialder[\lagr]{\bracket{\der[\vec{x}]{\sigma}}} = 0
\end{equation} 
or 
\begin{equation}
\label{24}
\vec{p} = - \frac{\vec{\pi}}{\pi_0}.
\end{equation} 
Eqs. \ref{20} and \ref{24} imply the constraint
\begin{equation}
\label{25}
 g^{ij}p_i p_j - 2g^{0i} p_i + g^{00} = 0.
\end{equation} 
Since the reparametrization invariance is the only gauge symmetry of the Lagrangian \ref{15}, eq. \ref{25} is the unique primary first class constraint. Its validity can be checked explicitly by computing $p_i$ from eq. \ref{15}. The Hamiltonian computed from $\lagr$ vanishes identically, $H \equiv 0$.

Now, let us impose the gauge condition \ref{17}. Together with eq. \ref{25} it forms the set of two second class constraints, so, in principle, they can be viewed as strong equalities provided we replace the Poisson bracket by Dirac one \cite{24}. However, the dynamics becomes trivial because the Hamiltonian vanishes in the strong sense as well. The reason is that the constraint \ref{17} is explicitly "time" dependent and Dirac's algorithm cannot be applied directly. To manage the problem we first perform the canonical transformation 
\begin{align}
&x'^a = x^a, \ a = 1, 2 
\\
&x'^3 = x^3 - \sigma 
\\
&p'_i = p_i, \ i = 1, 2, 3.
\end{align}
The relevant generating function \cite{25} reads 
\begin{equation}
\label{29}
\psi\bracket{\vec{x}, \vec{p}\ ', \sigma} = \sum\limits^2_{a=1} x^a p'_a + \bracket{x^3 - \sigma} p'_3
\end{equation}
leading to the new Hamiltonian
\begin{equation}
\label{30}
\ham' = \ham + \partialder[\psi]{\sigma} = \ham - p'_3.
\end{equation}  
Now, the constraint \ref{17} becomes time independent, $x'^3 = 0$. Together with eq. \ref{25} it can be used to eliminate $x^3$ and $p_3$ so we are left with canonical variables $x^a$, $p_a$, $a=1,2$. For the latter the Dirac bracket reduces to the Poisson one while, according to eq. \ref{30}, the new Hamiltonian, due to $\ham\equiv 0$, equals $-p_3$ (we omitted the already unnecessary prime). 

The constraint \ref{25}, which serves for computing $p_3$ can be rewritten as 
\begin{align}
\nonumber
F\bracket{x^1, x^2, p_1, p_2; \ham} = g^{ab} p_a p_b - 2g^{a3} p_a \ham + g^{33} \ham^2 +
\\
\label{31}
- 2g^{0a} p_a + 2g^{03} \ham + g^{00} = 0,
\end{align}
where the summation over $a, b$ runs from 1 to 2. 

By comparing eqs. \ref{2} and \ref{31} one finds the general form of the Hamiltonian $H$ equivalent to $\tilde{H} \equiv \ham$: 
\begin{align}
\nonumber
H = G\bracket{x^1, x^2, p_1, p_2} (g^{ab} p_a p_b - 2g^{a3} p_a \lambda + g^{33} \lambda^2 + 
\\
\label{32}
- 2g^{0a} p_a + 2g^{03} \lambda + g^{00}),
\end{align}
with arbitrary nonvanishing identically $G$. Let us denote by \textcolor{blue}{$\tau$} the evolution parameter entering the canonical equations defined by H. Then eq. \ref{7} yields 
\begin{equation}
\label{33}
\der[x^3]{\tau} = 2G\bracket{g^{a3}p_a - g^{03} - \lambda g^{33}} .
\end{equation}

\section{Kerr metric, Carter constant and integrability of Fermat's principle}

Let us apply now the results of previous sections to the ray trajectories in Kerr metric. In Boyer-Lindquist coordinates (with $x^0=t$, $x^1=r$, $x^2=\theta$, $x^3=\phi$) it reads:
\begin{equation}
\label{34}
\dt[s]^2 = g_{tt} \dt[t]^2 + g_{rr} \dt[r]^2 + g_{\theta \theta} \dt[\theta]^2 + g_{\phi \phi} \dt[\phi]^2 + 2 g_{t \phi} \dt[t] \dt[\phi],
\end{equation}
where $\bracket{\Delta \equiv r^2 - 2 m r + k^2, \rho^2 \equiv r^2 + k^2 \cos^2 \theta}$
\begin{align}
\label{35}
g_{tt} &= \frac{\Delta - k^2 \sin^2 \theta}{\rho^2}\\
g_{rr} &= - \frac{\rho^2}{\Delta}\\
g_{\theta \theta} &= - \rho^2\\
g_{\phi \phi} &= \frac{\sin^2 \theta \bracket{\Delta k^2 \sin^2 \theta - \bracket{r^2 + k^2}^2}}{\rho^2}\\
\label{39}
g_{t\phi} &= \frac{k \sin^2\theta\bracket{r^2 + k^2 - \Delta}}{\rho^2}
\end{align}
In order to compute the relevant Hamiltonian we use eq. \ref{32}. A convenient choice for $G$ is 
\begin{equation}
\label{40}
G = \frac{1}{2}g_{rr},
\end{equation}
which makes the coefficient in front of $p^2_r$ equal $\frac{1}{2}$: 
\begin{align}
H = \frac{1}{2} p^2_r + \frac{g_{rr}}{2 g_{\theta \theta}} p^2_\theta + \frac{g_{rr} \bracket{g_{tt} \lambda^2 - 2 g_{t \phi} \lambda}}{2 \bracket{ g_{tt} g_{\phi \phi} - g^2_{t\phi} } }
+ 
\frac{g_{rr} g_{\phi \phi}}{2 \bracket{ g_{tt} g_{\phi \phi} - g^2_{t\phi} } }
\end{align}
or, explicitly, 
\begin{align}
\label{42}
\nonumber
H = \bracket{ \frac{1}{2} p^2_r - \frac{\bracket{r^2 + k^2}^2}{2\Delta^2} - \frac{ 2mrk}{\Delta^2} \lambda - \frac{ k^2}{2 \Delta^2} \lambda^2}+
\\
 + \frac{1}{2\Delta} \bracket{ p^2_\theta + \frac{\lambda^2}{\sin^2 \theta} + k^2 \sin^2\theta}.
\end{align}

It is obvious from eq. \ref{42} that we are dealing with two-dimensional dynamical system which is integrable in Arnold--Liouville sense. The $\theta$-dependent integral of motion $p^2_\theta + \frac{\lambda^2}{\sin^2 \theta} + k^2 \sin^2 \theta$ is the counterpart of the Carter constant \cite{3,4}; one easily finds using eq. \ref{24} that our integral of motion is the ratio of the latter and $\pi^2_0$. 

The results of the sec. II imply that the ray trajectories in Kerr metric are determined by the equations
\begin{align}
\label{43}
&p^2_\theta + \frac{\lambda^2}{\sin^2 \theta} + k^2 \sin^2 \theta = C
\\
\label{44}
&\frac{1}{2} p_r^2 - \frac{\bracket{r^2 + k^2}^2}{2 \Delta^2} - \frac{2mrk\lambda}{\Delta^2} - \frac{k^2}{2 \Delta^2} \lambda^2 + \frac{C}{2 \Delta} = 0.
\end{align}
Together with the relation between the evolution parameter $\tau$ for the Hamilton equations determined by $H$ and the azimuthal angle $\phi$ (see below) eqs. \ref{43} and \ref{44} describe the shape of light rays. Note that the canonical equations of motion imply
\begin{align}
\label{45}
\der[r]{\tau} &= p_r
\\
\label{46}
\der[\theta]{\tau} &= \frac{p_\theta}{\Delta}.
\end{align} 
Expressing in \ref{43}, \ref{44} the momenta $p_\theta$ and $p_r$ in terms of the relevant derivatives one finds 
\begin{align}
\label{47}
&\Delta^2 \bracket{\der[\theta]{\tau}}^2 + \frac{\lambda^2}{\sin^2 \theta} + k^2 \sin^2 \theta =C
\\
\label{48} 
&\bracket{\der[r]{\tau}}^2 - \frac{\bracket{r^2 + k^2}^2}{\Delta^2} - \frac{4mrk\lambda}{\Delta^2} - \frac{k^2\lambda^2}{\Delta^2} + \frac{ C}{\Delta} = 0.
\end{align}
The above equations decouple if we introduce a counterpart of Mino time \cite{20}. Along any trajectory one defines a new evolution parameter $\color{blue}{\tilde{\tau}}$ through 
\begin{equation}
\label{49}
\der[\tilde{\tau}]{\tau} = \frac{\lambda}{\Delta}. 
\end{equation} 
Then eqs. \ref{47} and \ref{48} take the form 
\begin{align}
\label{50}
&\bracket{\der[\theta]{\tilde{\tau}}}^2 + \frac{1}{\sin^2\theta} + \frac{k^2}{\lambda^2} \sin^2\theta = \frac{C}{\lambda^2}
\\
\label{51}
&\bracket{\der[r]{\tilde{\tau}}}^2 - \frac{\bracket{r^2 + k^2}^2}{\lambda^2} - \frac{4mrk}{\lambda} - k^2 + \frac{C\Delta}{\lambda^2} = 0  
\end{align} 
There remains to determine the azimuthal angle $\phi$. To this end one can use eqs. \ref{33} and \ref{49}: 
\begin{equation}
\label{52}
\der[\phi]{\tilde{\tau}} = \frac{1}{\sin^2\theta} - \frac{k^2}{\Delta} - \frac{2mkr}{\lambda \Delta}
\end{equation}
Once $\theta = \theta\bracket{\tilde{\tau}}$, $r = r\bracket{\tilde{\tau}}$ are found from eqs. \ref{50} and \ref{51}, $\phi = \phi\bracket{\tilde{\tau}}$ is obtained by straightforward integration of \ref{52}. 

Equations \ref{50}, \ref{51} can be converted into nice ones for polynomial oscillators by making simple redefinitions of variables. First, under 
\begin{equation}
\label{53}
z \equiv \cos\theta 
\end{equation}
eq. \ref{50} becomes
\begin{equation}
\label{54}
\bracket{\der[z]{\tilde{\tau}}}^2 + \bracket{ \frac{C}{\lambda^2} - \frac{2k^2}{\lambda^2}} z^2 + \frac{k^2}{\lambda^2} z^4 = \frac{C}{\lambda^2} - \frac{k^2}{\lambda^2} - 1
\end{equation} 
which describes the motion of particle of unit mass in the simple quartic potential
\begin{equation}
\label{55}
U\bracket{z} = \bracket{\frac{C}{2\lambda^2} - \frac{ k^2}{\lambda^2}} z^2 + \frac{k^2}{2 \lambda^2} z^4 
\end{equation} 
with the energy $\frac{1}{2}\bracket{\frac{C}{\lambda^2} -\frac{k^2}{\lambda^2} - 1}$. Note that \ref{53} describes a one-to-one mapping from $\left<0, \pi\right>$ to $\left<-1, 1 \right>$.
\\
As far as eq. \ref{51} is concerned, one makes the standard substitution
\begin{equation}
\label{56}
u \equiv \frac{m}{r}.
\end{equation} 
Then eq. \ref{51} takes the form 
\begin{align}
\label{57}
\nonumber
\bracket{\der[u]{\tilde{\tau}}}^2 + \bracket{\frac{C}{\lambda^2} - \frac{2 k^2}{\lambda^2} } u^2 - 2 \bracket{  \frac{C}{\lambda^2} + \frac{2k}{\lambda} }u^3
\\
+ \frac{k^2}{m^2} \bracket{\frac{C}{\lambda^2} - 1 - \frac{k^2}{\lambda^2}} u^4 = \frac{m^2}{\lambda^2}.
\end{align}
Again, we are dealing with the motion of a particle in quartic potential
\begin{equation} 
\label{58}
V\bracket{u} = \bracket{\frac{C}{2\lambda^2} - \frac{k^2}{\lambda^2}} u^2 - \bracket{\frac{C}{\lambda^2} + \frac{2k}{\lambda}} u^3 + \frac{k^2}{2 m^2} \bracket{\frac{C}{\lambda^2} - 1 - \frac{k^2}{\lambda^2}} u^4 
\end{equation} 
carrying the energy $\frac{m^2}{2\lambda^2}$.
\\
Concluding, the light propagation in Kerr black hole can be described in terms of dynamics of two independent quartic oscillators describing the behaviour of $r$ and $\theta$ coordinates together with the equation
\begin{equation}
\label{59}
\der[\phi]{\tilde{\tau}} = \frac{1}{1-z^2} - \frac{\frac{k^2}{m^2} u^2}{1 - 2u + \frac{k^2}{m^2} u^2} - \frac{2 \bracket{\frac{k}{\lambda}} u}{1 - 2u + \frac{k^2}{m^2} u^2}.
\end{equation}
It follows that the relevant trajectories can be expressed in terms of elliptic functions.

\section{Weak deflection regime} 

The formalism described above is particularly suited to the description of light propagation in the asymptotic limit when the perihelion distance is large in comparison with the black hole radius. The former is basically determined by the value of invariant impact parameter defined as the ratio of conserved momenta, $\abs{\frac{\pi_3}{\pi_0}}$. According to the results of previous sections $\frac{\pi_3}{\pi_0} = \tilde{H} = \lambda$. Therefore, the weak deflection regime is defined by $|\lambda| \gg m$. Consequently, the small expansion parameter can be chosen to be $\frac{m}{\lambda}$. In order to make this expansion meaningful one has to specify the way we choose the initial conditions. Since the metric under consideration is axially symmetric one can assume that the perihelion is attained at $\phi=0$.
\\
Therefore, we put 
\begin{align}
\label{60}
&r\bracket{\phi=0} = r_{min} \hspace{5mm}  \bracket{u\bracket{\phi=0} = u_{max}}
\\
\label{61}
&\left. \der[r]{\phi} \right|_{\phi=0} = 0 \hspace{11mm}  \bracket{ \left.\der[u]{\phi}\right|_{\phi=0} = 0}
\end{align}  
It is easy to show (see below) that $r_{min}$ is determined by $\lambda$ and $r_{min} = O\bracket{\lambda}$. In order to deal with \ref{60} and \ref{61} in as simple way as possible, we impose the condition $ \phi\bracket{\tilde{\tau}=0} = 0$;
then, by eq. \ref{59}:
\begin{equation}
\label{62}
\phi = \int_0^{\tilde{\tau}} \left( \frac{1}{1 - z^2} - \frac{\frac{k^2}{m^2}  u^2}{1 - 2u + \frac{k^2}{m^2} u^2} -  \frac{2 \bracket{\frac{k}{\lambda}}u}{1 - 2u + \frac{k^2}{m^2} u^2} \right) d\tilde{\tau} 
\end{equation}
There remains to specify the initial conditions for $\theta$. The natural choice is to assume that $ \theta\bracket{\phi=0} 
$
and
$
    \frac{d\theta}{d\phi}\bracket{\phi=0} 
$
are $\lambda$-independent. However, our starting point is the more tractable eqs. \ref{54} and \ref{57}. Therefore, it is more convenient to impose the $\lambda$-independent initial conditions
\begin{equation}
\label{63}
    \theta\bracket{\tilde{\tau}=0} \equiv \theta_0 
\end{equation} 
\begin{equation}
    \left. \der[\theta]{\tilde{\tau}} \right|_{\tilde{\tau}=0} \equiv \dot{\theta}_0 
\end{equation} 
Then, due to $\phi\bracket{\tilde{\tau}=0} = 0$, we have $\theta\bracket{\phi=0} = \theta_0$; however,
$
    \left. \der[\theta]{\phi} \right|_{\phi=0} 
$
will no longer be $\lambda$-independent. This is not a serious obstacle since once we find the parametrized trajectory $r = r\bracket{\tilde{\tau}}$, $\theta = \theta(\tilde{\tau})$, $\phi = \phi(\tilde{\tau})$, given in terms of $\lambda$-independent initial conditions at $\tilde{\tau} = 0$, one can recompute it to a given order in $\frac{m}{\lambda}$, assuming definite $\lambda$-independent initial conditions at $\phi=0$. Putting $\tilde{\tau}=0$ in eq. \ref{50}, we find
\begin{equation}
\label{65}    
\dot{\theta}_0^2 + \frac{1}{\sin^2\theta_0} + \frac{k^2}{\lambda^2} \sin^2\theta_0 = \frac{C}{\lambda^2} \equiv C_0 + C_1 f^2,
\end{equation} 
where we have introduced the new small parameter
\begin{equation}
\label{66}
    f \equiv \frac{k}{\lambda} = \left( \frac{k}{m} \right) \left( \frac{m}{\lambda} \right).
\end{equation}
It follows from eq. \ref{65} that there are two $\lambda$-independent constants:
\begin{equation}
    C_0 \equiv \dot{\theta}_0^2 + \frac{1}{\sin^2\theta_0}, \quad C_0 > 1
\end{equation} 
\begin{equation}
    C_1 \equiv \sin^2\theta_0, \quad 0 \leq C_1 \leq 1.
\end{equation} 
Instead of $\frac{m}{\lambda}$, one can use $f$, eq. \ref{66}, as the small parameter describing the weak deflection regime. In order to recast appropriately eqs. \ref{54} and \ref{57}, we define
\begin{equation}
    y = fz
\end{equation}
and rewrite them in the form
\begin{equation}
\label{70}
    \left(  \der[y]{\tilde{\tau}} \right)^2 + (\alpha + \beta f^2) y^2 + y^4 = (\alpha - 1) f^2 + (\beta + 1) f^4
\end{equation}
\begin{align}
\nonumber
\left( \der[u]{\tilde{\tau}} \right)^2 + (\alpha + \beta f^2) u^2 - 2 \bracket{\alpha + 2f + (\beta + 2) f^2} u^3 + 
\\
\label{71}
+ \frac{k^2}{m^2} \left( (\alpha - 1) + (\beta + 1) f^2 \right) u^4 = \left( \frac{m}{k} \right)^2 f^2
\end{align} 
with  
\begin{align}
\label{72}
&\alpha \equiv C_0, \quad \beta \equiv C_1 - 2
\\
\nonumber
&\text{i.e.} \quad \alpha \geq 1, \quad -2 \leq \beta \leq -1.
\end{align}
For small $f$, eqs. \ref{70} and \ref{71} imply $y = O(f)$ and $u = O(f)$, respectively. This is obvious for $y$. On the other hand, $V(u)$, eq. \ref{58}, takes the form
\begin{align}
\nonumber
V(u) = \frac{1}{2} (\alpha + \beta f^2) u^2 - (\alpha + (\beta + 2) f^2 + 2f) u^3 +
\\
\label{73}
+ \frac{k^2}{2m^2} ((\alpha - 1) + (\beta + 1) f^2) u^4
\end{align} 
The shape of $V(u)$ is sketched on Fig. 1.

\begin{figure}[H]
\centering
\begin{tikzpicture}[scale=1]
\begin{scope}[xscale = 8,yscale = 30]
\draw[->] (-0.7, 0) -- (1, 0) node[right] {$u$};
\draw[->] (0, -0.2) -- (0, 0.2) node[above] {$V$};
\draw[domain=-0.3:0.7,smooth,variable=\x] plot ({\x},{0.5*(2-0.015)*\x*\x -(2+0.005 + 0.2)* \x*\x*\x + 0.45*(1-0.005)*\x*\x*\x*\x});
\draw[dotted] (0,1/29) node[left]  {$V_{max}$} -- (0.33, 1/29);
\end{scope}
\end{tikzpicture}
\caption{The shape of $V(u)$ for $\alpha=2$, $\beta=-1.5$ and $f=0.1$.} \label{rys1}
\end{figure}

%
It is important to note that $V_{max} = O(1)$. Therefore, eq. \ref{71} describes also small oscillations for sufficiently small $f$. One can apply well-developed techniques for solving nonlinear oscillating equations.
\\
Eqs. \ref{70} and \ref{71} imply the following equations of motion:
\begin{equation}
\label{74}
\sder[y]{\tilde{\tau}} + (\alpha + \beta f^2) y + 2 y^3 = 0
\end{equation} 
\begin{align}
\label{75}
\sder[u]{\tilde{\tau}} + (\alpha + \beta f^2) u - 3 (\alpha + 2f + (\beta + 2) f^2) u^2 +
\\
\nonumber
+ \frac{2k^2}{m^2} ((\alpha - 1) + (\beta + 1) f^2) u^3 = 0
\end{align}

The elegant and efficient method of providing the approximate solution to the equations describing small nonlinear oscillations has been proposed by Lindstedt and Poincaré \cite{21,22,23}. It consists in partial expansion in small amplitude of oscillations which, due to $y=O(f)$, $u=O(f)$, is in fact an expansion in $f$. However, we do not expand the arguments of trigonometric functions which also involve $f$ due to the fact that the frequency of nonlinear oscillations depends on amplitude. Instead, one determines the frequency order by order by demanding the absence of resonant terms; the application of the Lindstedt--Poincaré method to light propagation in the weak deflection regime is described in some detail in \cite{26}.

Once the equations \ref{74} and \ref{75} are solved to a given order, the arbitrary constants they involve are determined from initial conditions and/or the energy integrals \ref{70}, \ref{71}. Let us also note that the $\theta$ angle can be determined, using the Lindstedt--Poincaré idea, directly from the equation obtained by differentiating eq. \ref{54}. When applied to eq. \ref{74}, the Lindstedt--Poincaré method yields:
\begin{align}
\label{76}
& z(\tilde{\tau}) = A \cos \tau \color{black}{+ (B \cos \tau \color{black}{ + \frac{A^3}{16 \alpha} \cos 3 \tau\color{black}{) f^2 + O(f^4)}}}
\\
\label{77}
& \tau \color{black}{= \omega(\tilde{\tau} + \tilde{\tau}_0)}
\\
\label{78}
& \omega = \sqrt{\alpha} + \frac{1}{2\sqrt{\alpha}} \left( \beta + \frac{3}{2} \frac{ \alpha - 1 }{\alpha} \right) f^2 + O(f^4).
\end{align}
Inserting the above solution into the energy integral \ref{70} determines the constants $A$ and $B$:
\begin{align}
\label{79}
& \alpha A^2 = \alpha - 1
\\
\label{80}
& 2\alpha AB + \beta A^2 + \frac{9}{8} A^4 = \beta + 1
\end{align}
There remains one arbitrary constant $\tilde{\tau}_0$. The initial conditions \ref{62}, \ref{63} may be rewritten as:
\begin{align}
\label{81}
&z^2(0) = - (\beta + 1)
\\
\label{82}
&\dot{z}^2(0) = \alpha(\beta + 2) - 1
\end{align} 
However, we have already used the energy integral, so \ref{81} and \ref{82} should be equivalent. It is not difficult to verify that indeed this is the case. Putting
\begin{equation}
\label{83}
\cos^2 (\omega \tilde{\tau}_0) \equiv \mu + \nu f^2 + O(f^4), \quad \tau_0 \equiv \omega \tilde{\tau}_0
\end{equation}
one finds from \ref{81} (or \ref{82}):
\begin{align}
\label{84}
& \mu A^2 + \beta + 1 = 0
\\
\label{85}
& (2AB - \frac{3 A^4}{8 \alpha}) \mu + \frac{A^4}{2 \alpha} \mu^2 + \nu A^2 = 0
\end{align} 
To the second order in $f$, the behavior of the $\theta$ angle is determined by eqs. \ref{76}–\ref{78}, with the constants $A$, $B$ computed from \ref{79}, \ref{80} and $\tilde{\tau}_0$ given by \ref{83}, \ref{84} and \ref{85}. The behavior of the radial variable is given by eqs. \ref{75}, \ref{71} together with the boundary conditions \ref{60}, \ref{61}. By applying the Lindstedt--Poincaré algorithm, one obtains:
\begin{align}
\nonumber
& u(\tilde{\tau}) = \bar{A} f \cos (\bar{\omega} \tilde{\tau}) + \left( \frac{3}{2} - \frac{1}{2} \cos (2 \bar{\omega} \tilde{\tau}) \right) \bar{A}^2 f^2 +
\\
\label{86}
& + \left( 3 - \cos(2\bar{\omega} \tilde{\tau}) \right)\frac{f^3 \bar{A}^2}{\alpha} + \left( \frac{3}{16} + \frac{k^2 (\alpha-1)}{16 m^2 \alpha} \right) f^3 \bar{A}^3 \cos(3 \bar{\omega} \tilde{\tau}) +
\\
\nonumber
& + \bracket{\frac{37}{16} \bar{A}^3 - \frac{9 k^2 (\alpha - 1)}{16 m^2 \alpha} \bar{A}^3 - \frac{\beta}{2 \alpha} \bar{A}} f^3 \cos(\bar{\omega} \tilde{\tau}) + O(f^4)
\end{align}
\begin{equation}
\bar{\omega} = \sqrt{\alpha} + \frac{1}{2\sqrt{\alpha}} \left( \beta - \frac{15 \alpha}{2} \bar{A}^2 + \frac{3k^2 (\alpha -1)}{2 m^2} \bar{A}^2 \right) f^2 + O(f^4)
\end{equation} 
\begin{equation}
\label{88}
\bar{A} = \frac{m}{k \sqrt{\alpha}}
\end{equation}
This can be continued to higher orders, step by step, performing only algebraic operations. Once $\theta(\tilde{\tau})$ and $u(\tilde{\tau})$ are determined up to a given order, we recover $\phi(\tilde{\tau})$ to this order by simple integration (cf. eq. \ref{62}).
\\
Due to the fact that both $u$ and $\theta$ are given as linear combinations of trigonometric functions, it reduces again to the essentially algebraic operations.

It is interesting to see how the standard results for Schwarzschild metric are recovered. Putting $k=0$ in \ref{54}, \ref{70} and \ref{71} one finds
\begin{align}
\label{89}
&\bracket{\der[z]{\tilde{\tau}}}^2 + \alpha z^2 = \alpha-1
\\
\label{90}
&\bracket{\der[u]{\tilde{\tau}}}^2 + \alpha (u^2 - 2 u^3) = \frac{m^2}{\lambda^2}
\\
\label{91}
&\phi = \int_0^{\tilde{\tau}} \frac{\dt[\tilde{\tau}]'}{1 - z^2 (\tilde{\tau}')}
\end{align}   
Eqs. \ref{89} and \ref{91} can be immediately integrated leading to the relation
\begin{equation}
\label{92}
\sin \theta = \sqrt{\frac{1+\tan^2\phi}{1+ \alpha\tan^2\phi}}
\end{equation}
Eq. \ref{92} describes a plane. In particular, for $\alpha=1$ one obtains the 1-2 plane. Then eq. \ref{90} reduces to the standard one for the light rays in Schwarzschild metric. However, even if $\alpha>1$, one can easily check that standard description is recovered, provided one takes into account that the plane of motion is no longer the plane 1-2.

\section{5D example}

As it has been noticed in Sec. 3, the light propagation can be nicely described by Fermat's principle in Hamiltonian form provided:
(i) the gravitational field is time-independent;
(ii) at least one coordinate is cyclic.

Actually, the latter assumption is not crucial except that we prefer to work with autonomous systems. This makes the method proposed here quite flexible. As an example, let us apply it to the light propagation in Myers–Perry space-time \cite{27}, a higher-dimensional generalization of the Kerr space-time, which describes rotating black holes in more than four dimensions. \footnote{We thank an anonymous referee for suggesting this problem.}

The free motion in Myers–Perry space-times has been studied in \cite{28,29,30,31,32,33}. It is known that the relevant Hamilton–Jacobi equation is separable \cite{34,35,36,37}, which makes it possible to solve the geodesic equations in quadratures.

As an example showing that the problem of light propagation in Myers–Perry metric fits nicely into our scheme, we consider the five-dimensional Myers–Perry metric with a single rotation parameter. The relevant length element reads \cite{28}:

\begin{equation}
\label{95}
\begin{split}
\dt[s]^2 = \left(1 - \frac{\mu}{\rho^2}\right) \dt[t]^2
- \frac{\rho^2 r^2}{\Delta} \dt[r]^2 - \rho^2  \dt[\theta]^2
- r^2 \cos^2 \theta \dt[\psi]^2
\\
- \left((r^2 + a^2) + \frac{\mu a^2}{\rho^2} \sin^2 \theta \right) \sin^2\theta  \dt[\phi]^2 - \frac{2\mu a}{\rho^2} \sin^2\theta \dt[t]  \dt[\phi]
\end{split}
\end{equation}
where $\mu > 0$ and $a$ are constants related to mass and angular momentum of the black hole, respectively, while
\begin{equation}
\Delta \equiv r^2(r^2 + a^2) - \mu r^2
\end{equation}
and $\rho^2$, as previously, is given by
\begin{equation}
\rho^2 = r^2 + a^2 \cos^2\theta
\end{equation}
The metric \ref{95} is time-independent and the angles $\phi$ and $\psi$ are cyclic coordinates.
Any of them (or any combination thereof) can be chosen as the evolution parameter in our approach; in what follows, we choose $\psi$.

Following the prescription outlined in Sec. 3, we find the following equation for the Hamiltonian $\ham$ generating the evolution of light trajectories in the $\psi$ parameter:

\begin{equation}
\label{98}
g^{rr} p_r^2 +
g^{\theta\theta} p_\theta^2 +
g^{\phi\phi} p_\phi^2 +
g^{\psi\psi} \ham^2 -
2g^{t\phi} p_\phi +
g^{tt} = 0
\end{equation}
with
\begin{align}
\label{99}
g^{tt} &= - \frac{\bracket{(r^2 + a^2) \rho^2 + \mu a^2 \sin^2 \theta}}{(\mu - \rho^2) (r^2 + a^2) - \mu a^2 \sin^2 \theta}
\\
g^{\phi \phi} &= \frac{\rho^2 - \mu}{\bracket{(\mu - \rho^2) (r^2 + a^2) -\mu a^2 \sin^2 \theta} \sin^2 \theta}
\\
g^{t \phi} &= \frac{\mu a}{(\mu - \rho^2) (r^2 + a^2) -\mu a^2 \sin^2 \theta}
\\
g^{rr} &= - \frac{\Delta}{r^2 \rho^2}
\\
g^{\theta \theta} &= -\frac{1}{\rho^2}
\\
g^{\psi \psi} &= - \frac{1}{r^2 \cos^2 \theta}
\end{align}
According to the eq. \ref{34} the equivalent Hamiltonian reads
\begin{equation}
\label{105}
H = G \bracket{g^{rr} p_r^2 + g^{\theta \theta} p^2_\theta + g^{\phi \phi} p^2_\phi + g^{\psi \psi} \lambda^2  - 2 g^{t\phi} p_\phi + g^{tt}}
\end{equation}
with $G$ being an arbitrary nonvanishing function of $r$, $\theta$, $\phi$, $p_r$, $p_\theta$, $p_\phi$. Again, a convenient choice is $G = \frac{1}{2} g_{rr}$ yielding $\frac{1}{2}$ as the coefficient in front of $p_r^2$.
\\
Using eqs. \ref{99}-\ref{105} one finds
\begin{equation}
\label{106}
\begin{split}
H = \bracket{\frac{1}{2} p_r^2 - \frac{a^2 p_\phi^2 r^4}{2 \Delta^2} + \frac{a^2 \lambda^2}{2 \Delta} - \frac{\mu a p_\phi r^4}{\Delta^2} - \frac{(\Delta + \mu (r^2 + a^2)) r^4}{2 \Delta^2} }
\\
 + \frac{r^2}{2 \Delta} \bracket{p^2_\theta + \frac{p^2_\phi}{\sin^2 \theta} + \frac{\lambda^2}{\cos^2 \theta} - a^2 \cos^2 \theta}
\end{split}
\end{equation}
It follows immediately from eq. \ref{106} that the resulting dynamics is integrable. First, we find the counterpart of Carter's constant:
\begin{equation}
\label{107}
p_\theta^2 + \frac{p^2\phi}{\sin^2 \theta} + \frac{\lambda^2}{\cos^2\theta} - a^2 \cos^2 \theta = C
\end{equation}
On the other hand, keeping in mind that we are considering the submanifold $H=0$ (cf. Sec. 2), one obtains:
\begin{equation}
\label{108}
\frac{1}{2} p_r^2 - \frac{a^2 p_\phi^2 r^4}{2 \Delta^2} + \frac{a^2 \lambda^2}{2 \Delta} + \frac{\mu a p_\phi r^4}{\Delta^2} - \frac{(\Delta + \mu (r^2 + a^2)) r^4}{2 \Delta^2} + \frac{C r^2}{2 \Delta} = 0
\end{equation}
Denoting by $\tau$ the evolution parameter corresponding to the Hamiltonian $H$, we get the counterparts of eqs. \ref{45} and \ref{46}:
\begin{align}
\der[r]{\tau} &= p_r
\\
\der[\theta]{\tau} &= \frac{r^2}{\Delta} p_\theta
\end{align}
In terms of Mino-like time $\tilde{\tau}$, defined by the counterpart of eq. \ref{49},
\begin{equation}
\label{111}
\der[\tilde{\tau}]{\tau} = \frac{r^2}{\Delta}
\end{equation}
eqs. \ref{107} and \ref{108} reduce to:
\begin{align}
\label{112}
&\bracket{\der[\theta]{\tilde{\tau}}}^2 + \frac{p^2_\phi}{\sin^2 \theta} + \frac{\lambda^2}{\cos^2 \theta} - a^2 \cos \theta = C
\\
&\frac{1}{2} \bracket{\der[r]{\tilde{\tau}}}^2 - \frac{a^2 p^2_\phi}{2} + \frac{a^2 \lambda^2 \Delta}{2 r^4} - \mu a p_\phi - \frac{(\Delta + \mu (r^2+a^2) )}{2} + \frac{\Delta C}{2 r^2} = 0
\end{align}
Finally, the Hamiltonian equation for $\phi$ reads
\begin{equation}
\label{114}
\der[\phi]{\tilde{\tau}} = (a^2 p_\phi	+ \mu a) \frac{r^2}{\Delta}
\end{equation}
Eqs. \ref{112}--\ref{114} depend on three constants $\lambda$, $p_\phi$ and $C$. The first two can be solved by separating the variables and then $\phi(\tilde{\tau})$ is obtained from eq. \ref{114} by direct integration. This yields three additional constants. 

It remains to find $\psi$. Eq. \ref{7} implies
\begin{equation}
\label{115}
\der[\psi]{\tau} = \frac{\lambda\rho^2}{\Delta \cos^2 \theta}
\end{equation}
Combining \ref{111} and \ref{115} we finally obtain
\begin{equation}
\label{116}
\der[\psi]{\tilde{\tau}} = \frac{\lambda \rho^2}{r^2 \cos^2 \theta}
\end{equation}
which allows to find $\psi$, again by direct integration.

The above reasoning may be generalized to the case of both non-vanishing angular momenta. 
It is technically slightly more complicated because, in order to find the contravariant metric tensor, one has to compute the inverse of 3x3 matrix. Then the separability condition (the existence of Carter's constant counterpart) reads
\begin{equation}
g_{rr} g^{xy} = A^{xy}(r) + g_{rr} g^{\theta \theta} B^{xy} (\theta)
\end{equation}
for $x$,$y$ = $t$, $\phi$, $\psi$, and can be checked explicitly. 

\section{Final remarks}
The geodesics in curved space-time can be described in terms of Euler--Lagrange equations following from the homogeneous Lagrangian quadratic in velocities, provided the affine parameter is the evolution parameter. In the constant axially symmetric metric these Euler--Lagrange equations admit immediately three integrals of motion: the Lagrangian (numerically coinciding with the Hamiltonian) itself and the generalized momenta conjugated to $t$ and $\phi$. In order to provide the integrability in Arnold--Liouville sense one needs still one integral. It has been found by Carter \cite{3,4} and appears to be quadratic in generalized momenta. Therefore, it results from the symmetry described by canonical transformations, which do not reduce to the point ones. With four independent integrals of motion in involution the geodesic equations can be integrated by quadratures. In the case of Kerr's metric this results in description of geodesics in terms of elliptic functions (see, for example, \cite{6,7,8} and references therein). 

The relevant integrals of motion do not have a direct physical interpretation since they depend on the choice of affine parameter, which is defined up to an affine transformation. 

For the constant gravitational field, the alternative description within Lagrangian formalism is provided by the Fermat principle \cite{14} (for general gravitational field the Lagrange formalism must be replaced by the Herglotz variational principle \cite{19}). Then the time coordinate can be eliminated (it plays the role of action variable) and the resulting Lagrangian is reparametrization invariant. This allows to choose (at least locally) one of the coordinates as the evolution parameter. This is particularly convenient if the choice concerns cyclic coordinate. Then the dynamics is reduced to two-dimensional one. 

However, there is some price to be paid for this reduction. The Lagrangian describing Fermat's principle results from the solution of quadratic equation and is quite complicated. The same concerns the Hamiltonian. In order to simplify the formalism, we applied here the coupling constant metamorphosis method. It allowed us to replace the initial Hamiltonian by the family of simpler ones, parametrized by the initial "energy". As a result, we obtained the (family of) two-dimensional Hamiltonian system with the Hamiltonian quadratic in the momenta. In the case of Kerr's metric, the form of the Hamiltonian shows that it admits additional integral of motion (the counterpart of Carter’s constant) and, therefore, the dynamics is completely integrable. Moreover, after further redefinition of the evolution parameter (some kind of Mino time \cite{20}) it reduces to the system of two non-linear oscillators. Its form is suitable for perturbative calculations. They can be performed using efficient Lindstedt--Poincar\'e method \cite{21, 22, 23, 26}, providing successive terms of approximation to the Fourier expansion. 
We have also shown that the method proposed in the paper is also applicable to the problem of light propagation in Myers-Perry metric providing a higher dimensional generalization of the Kerr's metric. We considered 5d rotating black hole with a single rotation parameter and sketched the extension to the general case with two rotation parameters. It appeared that the coupling constant metamorphosis works here with no essential modification needed.
 
\section*{Acknowledgment} 
The fruitful discussions with Prof. K. Andrzejewski and Prof. P. Maslanka are gratefully acknowledged.

\bibliographystyle{ieeetr}
\bibliography{Kerr_metric}

\end{document}